\documentclass[sigconf]{acmart}

\AtBeginDocument{%
  \providecommand\BibTeX{{%
    Bib\TeX}}}

\copyrightyear{2023}
\acmYear{2023}
\setcopyright{acmcopyright}\acmConference[ASPDAC '23]{28th Asia and South Pacific Design Automation Conference}{January 16--19, 2023}{Tokyo, Japan}
\acmBooktitle{28th Asia and South Pacific Design Automation Conference (ASPDAC '23), January 16--19, 2023, Tokyo, Japan}
\acmPrice{15.00}
\acmDOI{10.1145/3566097.3567872}
\acmISBN{978-1-4503-9783-4/23/01}

\usepackage{multirow}
\usepackage{soul}
\usepackage[skip=0.5\baselineskip]{caption}

\usepackage{tikz}

\usepackage{amsmath,amsfonts}
\usepackage{algorithm,algorithmic}
\usepackage{graphicx}
\usepackage{textcomp}
\usepackage{xcolor}
\def\BibTeX{{\rm B\kern-.05em{\sc i\kern-.025em b}\kern-.08em
    T\kern-.1667em\lower.7ex\hbox{E}\kern-.125emX}}

\newcommand{\MVU}{MVU}
\newcommand{\MatrixVectorUnit}{Matrix Vector Unit}
\newcommand{\barvinn}{BARVINN}
\newcommand{\BARVINN}{BARVINN}

\newcommand{\Pito}{Pito: RISC-V-based Controller}
\newcommand{\pito}{Pito}
\newcommand{\barrel}{barrel}

\usepackage{color, colortbl}
\definecolor{ao}{rgb}{0.0, 0.5, 0.0}
\definecolor{amber}{rgb}{1.0, 0.75, 0.0}

\begin{document}

\title{BARVINN: Arbitrary Precision DNN Accelerator Controlled by a RISC-V CPU}


 \author{Mohammadhossein Askarihemmat\textsuperscript{1}, Sean Wagner\textsuperscript{2}, Olexa Bilaniuk\textsuperscript{3}, \\ Yassine Hariri\textsuperscript{4}, Yvon Savaria\textsuperscript{1}, Jean-Pierre David\textsuperscript{1}}
 \affiliation{%
  \institution{\textsuperscript{1}Ecole Polytechnique Montreal \country{Canada},
  \textsuperscript{2} IBM \city{Toronto}\country{Canada},
  \textsuperscript{3} Mila \city{Montreal}\country{Canada},
  \newline
  \textsuperscript{4} CMC Microsystems \city{Kingston}\country{Canada}}
{\{mohammad.hossein.askari-hemmat, yvon.savaria, jpdavid\}}@polymtl.ca,  wagnerse@ca.ibm.com, olexa.bilaniuk@mila.quebec, hariri@cmc.ca
}
    





\renewcommand{\shortauthors}{AskariHemmat et al}

\begin{abstract}
We present a DNN accelerator that allows inference at arbitrary precision with dedicated processing elements that are configurable at the bit level. Our DNN accelerator has 8 Processing Elements controlled by a RISC-V controller with a combined 8.2 TMACs of computational power when implemented with the  recent Alveo U250 FPGA platform. We develop a code generator tool that ingests CNN models in ONNX format and generates an executable command stream for the RISC-V controller. We demonstrate the scalable throughput of our accelerator by running different DNN kernels and models when different quantization levels are selected. Compared to other low precision accelerators, our accelerator provides run time programmability without hardware reconfiguration and can accelerate DNNs with multiple quantization levels, regardless of the target FPGA size. BARVINN is an open source project and it is available at \href{https://github.com/hossein1387/BARVINN}{https://github.com/hossein1387/BARVINN}.
\end{abstract}

\keywords{neural networks, hardware acceleration, FPGA, low-precision}

\maketitle

\section{Introduction}
Deep neural networks (DNNs) traditionally rely on floating point computations. These operations are slow and costly in terms of power consumption and required silicon area compared to fixed-point/integer operations. One way to accelerate computation in a DNN is to use less precision for computation via quantization \cite{DBLP:journals/corr/HubaraCSEB16}. This also reduces memory consumption as well as energy consumption. For instance, in a 45~nm process, 
8-bit integer multiplication and addition take 0.2~pJ and 0.03~pJ, respectively, while the same operations with 32-bit floating-point values requires 3.7~pJ for multiplication and 0.9~pJ for addition \cite{IEEE:ISSCC}. On an Intel Core i7 4770 running at 3.4GHz, multiplication is more than 3 times faster for fixed-point compared to floating-point \cite{LIMARE}.
With recent quantization techniques, these benefits are available with little to no loss in model performance and accuracy. In \cite{Esser2020LEARNED,lee2021network}, their quantization schemes showed accuracy losses of 1-3\% at 2-bit precision on most classification and object detection models.
Table \ref{tab:lsq_quant} illustrates the result of applying Learned Scale Quantization (LSQ) \cite{Esser2020LEARNED} with different bit precisions on different models and tasks. Quantized models offer accuracy similar to full precision models, while having smaller size.

\begin{table}[]
\centering

\caption{Effects of Quantization on Accuracy and Model Size.}

\begin{tabular}{|l|c|l|l|c|c|}
\hline
\multicolumn{1}{|c|}{Task} & Dataset & \multicolumn{1}{c|}{Model} & \multicolumn{1}{c|}{\begin{tabular}[c]{@{}c@{}}Precision\\ A/W\end{tabular}} & \begin{tabular}[c]{@{}c@{}}Acc/\\ MAP\end{tabular} & \begin{tabular}[c]{@{}c@{}}Size\\ (MB)\end{tabular} \\ \hline
\multirow{4}{*}{Classification} & \multirow{4}{*}{\begin{tabular}[c]{@{}c@{}}CIFAR \\ 100\end{tabular}} & \multirow{4}{*}{ResNet18} & LSQ(2/2) & 76.81 & 2.889 \\ \cline{4-6} 
 &  &  & LSQ(4/4) & 76.92 & 5.559 \\ \cline{4-6} 
 &  &  & LSQ(8/8) & 78.45 & 10.87 \\ \cline{4-6} 
 &  &  & FP32 & 76.82 & 42.8 \\ \hline
\multirow{4}{*}{\begin{tabular}[c]{@{}l@{}}Object \\ Detection\end{tabular}} & \multirow{4}{*}{\begin{tabular}[c]{@{}c@{}}VOC-\\ 2007\end{tabular}} & \multirow{4}{*}{\begin{tabular}[c]{@{}l@{}}SSD300-\\ ResNet18\end{tabular}} & LSQ(2/2) & 0.61 & 10.34 \\ \cline{4-6} 
 &  &  & LSQ(4/4) & 0.60 & 11.81 \\ \cline{4-6} 
 &  &  & LSQ(8/8) & 0.68 & 14.77 \\ \cline{4-6} 
 &  &  & FP32 & 0.59 & 32.49 \\ \hline
\end{tabular}
\label{tab:lsq_quant}
\vspace{-1cm}
\end{table}

Mixed-precision quantization \cite{micikevicius2017mixed,8954415,DBLP:conf/iclr/UhlichMCYGTKN20,DBLP:conf/aaai/YuLSH021,Bulat_2021_ICCV} further provides finer control to reach an optimal solution by learning different precisions for each layer of a network. In \cite{8954415}, the authors illustrate that using their mixed-precision framework, they reduced model latency and energy consumption by a factor of almost 2$\times$ with little drop in accuracy compared with an 8-bit quantized model. 

Fully benefiting from low-precision in a DNN requires hardware that natively supports low-precision computations. Commodity hardware can  perform arbitrary precision arithmetic by transforming data-layout and computing with bit-wise instructions \cite{10.1145/3368826.3377912}. However, this approach is extremely costly for general processors, because of the overhead for shifting, masking and packing bits to the correct format. At the time of writing this paper, there are no commercially available general processors (CPU or GPU) that can efficiently process data in arbitrary precision.  

In this paper, we propose an arbitrary low-precision DNN hardware accelerator called \BARVINN{}. 
Our accelerator is software programmable and can be integrated in the RISC-V standard development flow. It is designed as a highly optimized computational pipeline for DNNs that introduces low hardware overhead and offers low-power operation.
The contributions of our paper are as follows: 

\begin{itemize}
    \item Implementation of a DNN hardware accelerator with arbitrary fixed-point low-precision for matrix-vector multiply operations at high-throughput and low power.
    \item Implementation of a custom embedded RISC-V CPU to control an array of DNN accelerators by software.
    \item Data structures for efficiently storing and processing weights and activations for high-throughput serial computation.
    \item Development of a software code generator for transforming DNNs into RISC-V code that executes on our accelerator.

\end{itemize}

In section \ref{sec:related_works}, we review relevant DNN  accelerators from the literature. Section \ref{sec:design} presents the architecture of \BARVINN{}. In section \ref{sec:perf_analysis}, a detailed performance analysis of \BARVINN{} is provided and compared with other DNN accelerators. 

\section{Related works}
\label{sec:related_works}

Several DNN hardware accelerators supporting quantization and low-precision have been presented in recent years for both FPGA and ASIC targets. Here, we discuss the accelerator architectures most relevant to our work.


Recent FPGA-based accelerators include FINN \cite{umuroglu2017finn, blott2018finnr}, DNNBuilder \cite{zhang2018dnnbuilder}, and FILM-QNN \cite{sun2022film}. In FINN and DNNBuilder, a software toolchain is used to map a trained DNN to generated logic modules that are integrated together. An overall processing pipeline is generated and then synthesized for the target device. The advantage of this approach is that the logic efficiently implements a specific DNN with minimal overhead on a device that can be reconfigured to different DNNs at different times. However, this approach requires that all DNN layers be implemented in the logic all at once, which limits the size of the DNN to the amount of logic resources available on a given FPGA. While FINN supports low-precision down to binary and DNNBuilder down to 4-bit, neither supports arbitrary and mixed-precision at different DNN layers. In contrast, FILM-QNN does support DNN models of arbitrary sizes and quantized DNNs with mixed precisions. However, it is limited to only 4- or 8-bit weights and 5-bit activations due to a bit-packing scheme used with the DSP blocks in the FPGA.

Several ASIC accelerator designs support arbitrary precision. Bit Fusion \cite{sharma2018bit} uses a large array of 2-bit processing elements that can be fused together to perform up to 8-bit operations. Loom \cite{sharify2018loom}, and BitBlade \cite{ryu2019bitblade} employ bit serial computation schemes for added flexibility.
While bit-serial computation of any single math computation (e.g. multiplication) inherently requires additional clock cycles and latency over bit-parallel circuits, these designs exploit the large number of computations in a DNN that can be done in parallel. This is done by implementing a large number of bit-serial computational units operating simultaneously, which  compensates high latency with high throughput. For example, the Loom engine consists of $128 \times 16 = 2048$ Serial Inner-Product units (SIPs), each of which performs 16 $1 \times 1$-bit products per cycle.

\section{Architecture}
\label{sec:design}
\begin{figure*}[!h]
  \centering
    \includegraphics[width=\textwidth, height=10.89cm]{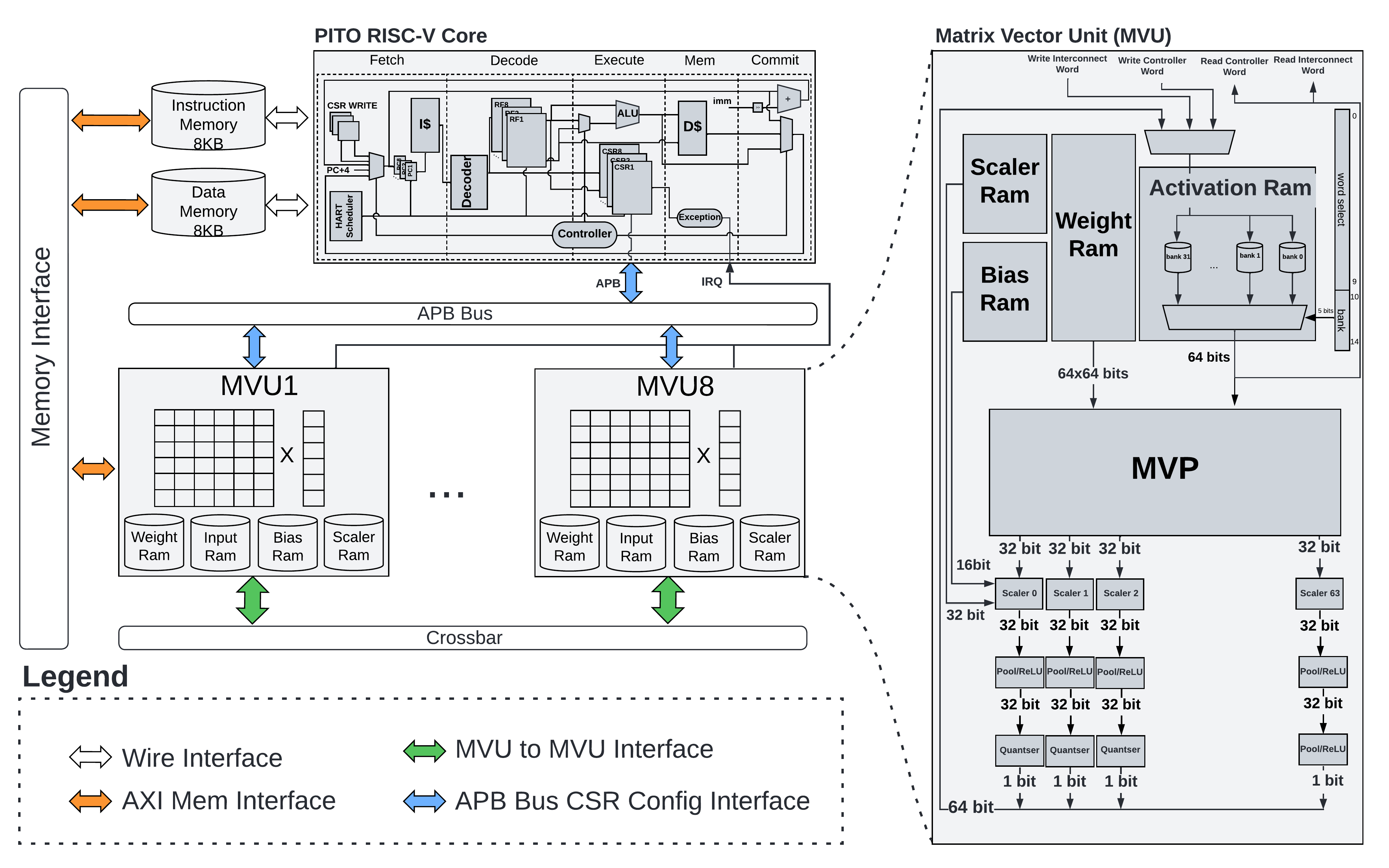}
    \small
\caption{\BARVINN{} hardware architecture with MVU array and Pito RISC-V controller. Right side is MVU detail.}
\label{fig:Barvinn_top}
\vspace{-2mm}
\end{figure*}

\BARVINN{} is designed to provide high-throughput and software programmability, while supporting DNNs of arbitrary size and type.
The high-level architecture of BARVINN is illustrated in Figure \ref{fig:Barvinn_top}.
It consists of the following main components: 1) an array of \MatrixVectorUnit{}s (\MVU{}) \cite{bilaniuk2019bitslice}, and 2) a RISC-V CPU
called \pito{} \cite{9401617}
as a controller for the \MVU{} array. The \MVU{}s accelerate common DNN computations such as GEMV, GEMM, and convolutions along with other operations such as batch normalization, ReLU activation, and quantization. \pito{} coordinates the computations in the \MVU{} array while also handling data transfers to and from the host system. 

As it is not possible to foresee all possible neural networks that may crop up in the literature in the future, high-level sequencing of tensor operations for BARVINN is done in software. 
To control the array of processing elements, unlike the aforementioned accelerators, BARVINN uses the standard RISC-V RV32I ISA. This allows us to leverage the pre-existing software ecosystem. Furthermore, by using a CPU that supports a well known ISA, BARVINN is more flexible and it can be adapted to support new DNN architectures.



\subsection{\MatrixVectorUnit{}s}
The base configuration of \barvinn{} is implemented with an array of 8 \MVU{}s. Figure \ref{fig:Barvinn_top} shows each \MVU{} is a 64-element vector pipeline with several modules: a) a Matrix Vector Product unit (MVP), b) RAMs for activations/weights/scalers/biases, c) a scaler unit, d) a pooling/activation unit, and e) a quantizer. \MVU{}s compute 64 output vector elements per clock cycle using a 64 element input data vector from the activation RAM and a 64$\times$64 element matrix from the weight RAM. Activation and weight RAMs store data in low-precision. MVP units operate in low-precision, while subsequent units in the pipeline operate in high-precision fixed-point.



\begin{figure}[t]
        \centering
            \includegraphics[width=8.0cm]{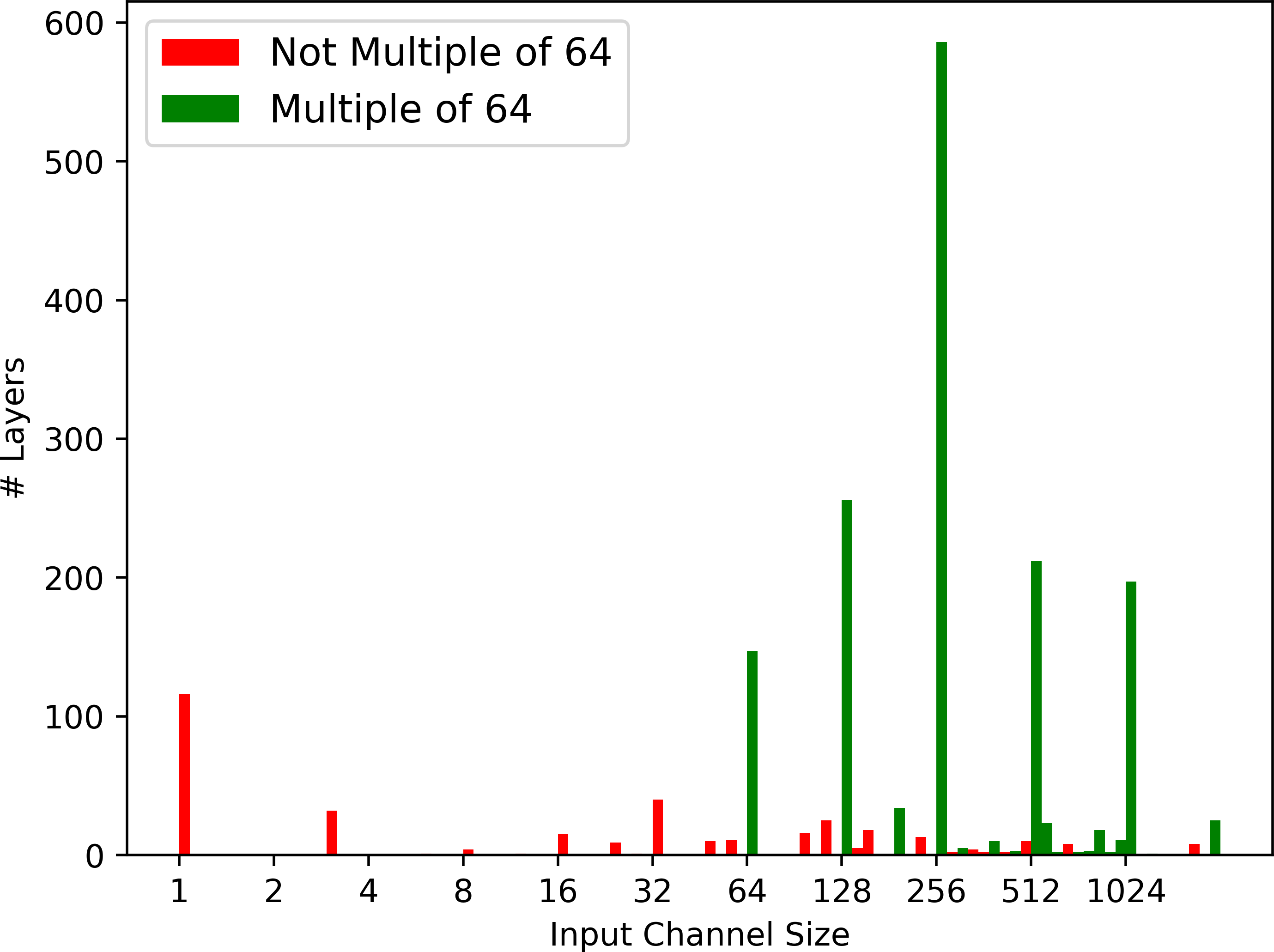}
        \caption{Channel sizes in models from ONNX Model Zoo.}
        \label{fig:ONNX_ZOO_ANALYSIS}
        \vspace{-2.5mm}
\end{figure}

To justify our design choice of operating on 64 element vectors, we analysed over 50 models available at the ONNX Model Zoo \cite{onnx_zoo} to check the input channel size of convolution layers. Figure \ref{fig:ONNX_ZOO_ANALYSIS} illustrates a distribution of input channel size of all layers among those models. We found that 79\% of these models use convolution with input channel sizes that are multiples of 64. 

\subsubsection{Matrix-Vector Product Units}

\begin{figure}[t]
    \hspace*{-1.3cm}  
      \centering
        \includegraphics[width=7.0cm]{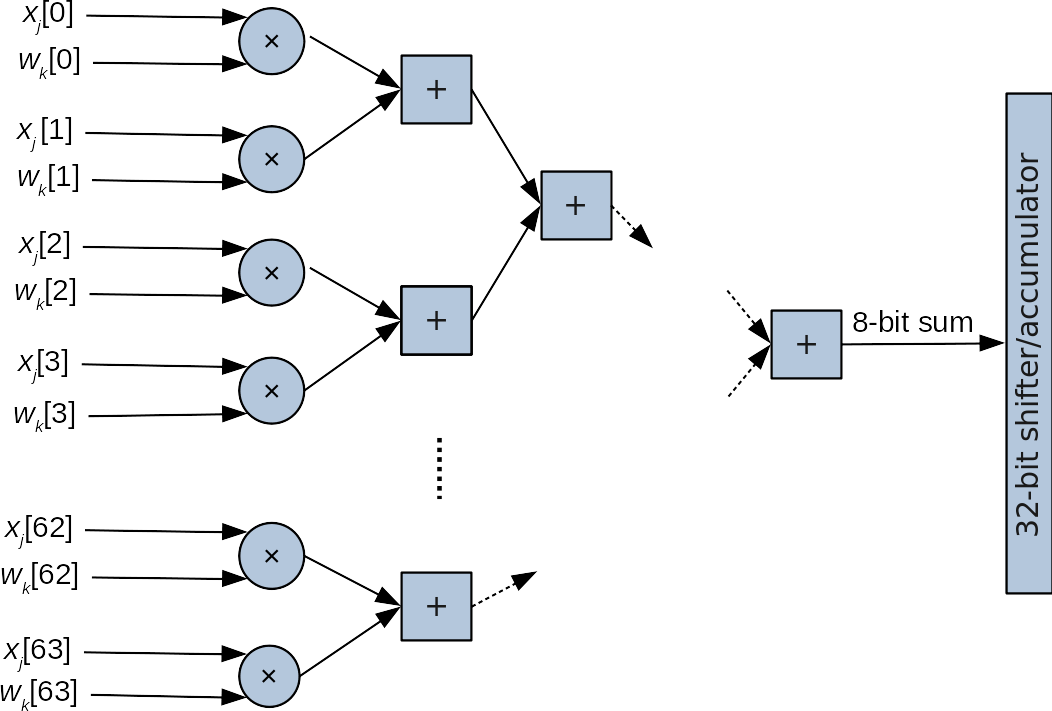}
        \small
    \caption{VVP unit with a shifter-accumulator. Bit $j$ from 64 elements of the activation tensor $x$ and bit $k$ from 64 elements of the weight tensor $w$ are input in a bit-serial fashion. Note that some input bits and layers of the 5-deep adder tree are not shown.}
    \label{fig:vvp_arch}
    \vspace{-4mm}
\end{figure}

Matrix operations are carried out by the MVP units. They compute on fixed-point arbitrary precision operands from 1- to 16-bit. Each MVP has 64 vector-vector product (VVP) pipelines. Each VVP has 64 input lanes with 1-bit multipliers, followed by an addition tree with 8-bit output, as shown in Figure \ref{fig:vvp_arch}. 
On every cycle, 64 bits from the activation RAM are broadcasted to each of the 64 VVPs, while a 64$\times$64 matrix tile from the weight RAM is read with each row of the tile sent to separate VVPs. The VVPs compute a 64-element dot product on 1-bit operands in each pipeline. With 64 VVPs per MVP, the overall output is a 64-element vector. 


MVPs compute arbitrary bit precision dot-products using the bit-serial scheme of \cite{bilaniuk2019bitslice}. Weights and activations can be unsigned or 2's-complement signed fixed-point. Bit-depth is set independently for both, thus allowing for mixed precision. The bit-serial dot-product, shown in Algorithm \ref{algo:bitserial}, is a multi-cycle sequence starting with the most significant bits (MSB) from 64 elements of the activation and weight tensors. Bits are multiplied in each lane and results are added together across lanes in an addition tree producing an 8-bit dot product. This is added to an accumulator/shifter (see Figure \ref{fig:vvp_arch}). The MSB$\times$MSB result represents the highest order-of-magnitude partial sum of the overall dot product. The MVP then computes the next lower order-of-magnitude partial sum by drawing the needed bit combinations of the operands. When a change in the order-of-magnitude is made, the accumulator is shifted left by 1-bit to align to the order-of-magnitude prior to adding the addition tree output. MVPs are fully pipelined, allowing them to work on different bit combinations at different stages without stalling. The operation completes when the dot products of the least significant bits (LSB) of the operands are computed and accumulated. For $b_w$-bit weights and $b_a$-bit activations, the overall operation takes $b_w b_a$ cycles to compute one tile of the output vector. The precision of the operands is configured separately for each \MVU{}, thus each \MVU{} can process different layers with different bit precisions.

\begin{algorithm}
\caption{Bit-serial dot-product}\label{algo:bitserial}
\begin{algorithmic}[1]
\STATE $b_a$, $b_w$: activation and weight bit precisions
\STATE $x, w$: activation and weight vectors of size $n$
\STATE $j, k$: bit position for activations and weights
\STATE $accumulator \gets 0$
\FOR{$i \gets b_w + b_a$ to $1$}
    \FORALL{$(j,k)$ where $j+k == i$}
        \FOR{$l \gets 0$ to $n-1$}
            \STATE $onebitprod = x_j[l] \times w_k[l]$
            \STATE $accumulator \gets accumulator + onebitprod$
        \ENDFOR
    \ENDFOR
    \item shift $accumulator$ left 1-bit
\ENDFOR
\STATE $output \gets accumulator$
\end{algorithmic}
\end{algorithm}

Our bit-serial dot product scheme differs from other architectures. The computation scheme in BitFusion is based on computing the individual products of the overall dot-product, that are then summed. This requires a large number of shift-registers to align and sum partial products. \BARVINN{} and BitBlade instead interchange the ordering of the computation such that partial products of the same magnitude from all individual products are computed first and then summed. This reduces the number shifters needed. \BARVINN{} additionally serializes the computation of partial products of different magnitude, requiring only a single fixed shifter and a single adder tree, whereas BitBlade requires 16 variable shifters and 17 adder trees. \BARVINN{} maintains throughput despite this serialized scheme by parallelizing across a wider number of input operands and producing a larger number of output products per clock cycle. BitFusion and BitBlade are further limited to operand sizes 2, 4, and 8-bit, whereas \MVU s in \BARVINN{} and SIPs in Loom support operands of any bit-depth down to 1-bit. However, Loom's data loading scheme restricts the efficiency for general matrix multiply operations when the weight bit depth is below 16, whereas \BARVINN{}s is able to maintain full throughput down to 1-bit.

\subsubsection{Memories and Data Layout}\label{memorylayout}
\label{subsec:mem_layout}
\begin{figure}[!h]
  \centering
    \includegraphics[width=8.0cm]{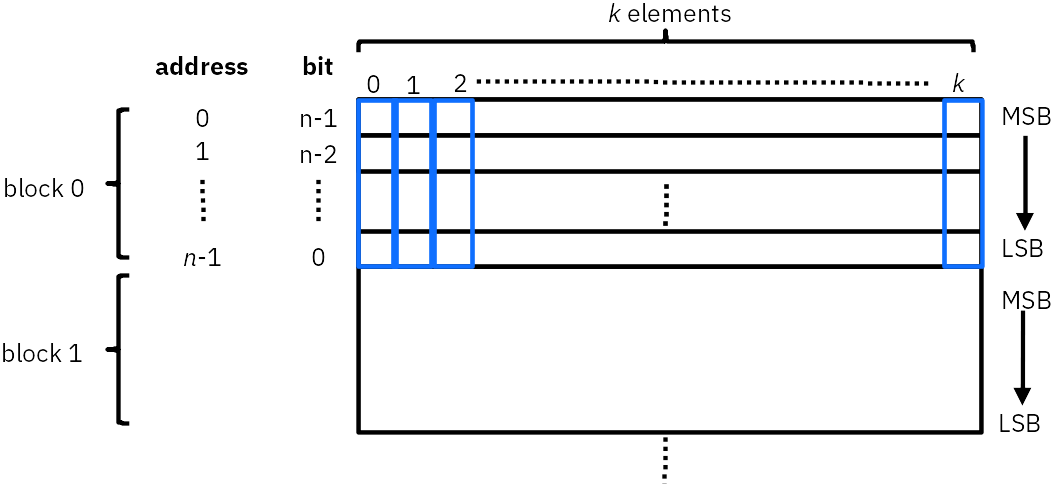}
    \small
\caption{Bit-transposed data format for arbitrary precision.}
\label{fig:bit-trans_mem_layout}
\vspace{-2mm}
\end{figure}

Activation and weight RAMs store data in a bit-transposed format shown in Figure \ref{fig:bit-trans_mem_layout} to exploit bit-serial computation. When precision is greater than 1 bit, tensor elements are organized in blocks where bits of the same order-of-magnitude are stored in the same memory word starting with the MSBs in the lowest address. A block of $n$ elements with precision $b$ requires $b$ memory words of width $n$. Activation vector elements are in blocks of 64
while weights matrix elements are in blocks of 4096 bits in order to load a 64$\times$64 matrix tile.
A transposer module transforms input data from the host into the needed bit-transposed format. Transposition is only needed on the first layer of a DNN since \MVU{}s write back to activation RAM in the bit-transposed format. Weights are pre-processed by a toolchain on the host and loaded into weights RAMs in the expected bit-transposed format.

The layout of the tensors in the RAMs depends on the operation to be performed. For GEMV, activations are organized as vectors with blocks of 64 elements, while weight matrices are organized as a set of 64$\times$64 tiles. For 2D convolutions, layout of activations is $NHWC$, where the channel dimension $C$ is the innermost dimension, followed by width $W$, height $H$, and batch size $N$. The $C$ dimension is the innermost since several common DNNs such as ResNet typically have hidden layer channel depths that are powers of 2, and hence align to the 64 input lanes of the VVPs. When there are more than 64 channels, the first 64 channels are stored in the first block, the second 64 channels are stored into the second block and so on. As an example, an input tensor of [N=1, H=8, W=8, C=256] with 2-bit precision, will have 4 channel blocks, each block will have 64 rows of 2 by 64-bit elements.

Our weight tensor memory layout for 2D convolutions is designed to support efficient execution by interleaving the input channel dimension $C_i$ and output channel dimension $C_o$. Each weight memory word contains 64 subsets from the $C_o$ dimension, with each subset containing 64 elements from the $C_i$ dimension. A contiguous block of $b_w$ words that stores a complete set of bits for the needed weight precision is referred to as a channel block $C_b$. The layout for 2D convolution weights is $C_{o,s}F_HF_WC_b$, where $C_{o,s} = C_o/64$ are output channel sets, and the kernel size is $(F_W,F_H)$. 

 

\subsubsection{Job Configuration and Execution}

\MVU{}s are programmed to perform jobs such as GEMV or Conv2D operation. A controller sets configuration registers that orchestrate the sequence of calculations and memory reads to complete an operation in the \MVU{}s. 
Once the job is finished, the \MVU{} will generate an interrupt to the controller, indicating that the job is finished and results are ready to be sent back to the host or to trigger subsequent operations on the same \MVU{} or other \MVU{}s. While a \MVU{} is busy, it can be programmed to prepare the next job to minimize idle time. 


Each \MVU{} contains address generation units (AGU) that drive the memory access pattern across the activation and weight RAMs. 
The access pattern is managed by a set of up to five nested loops with parameters setting the number of iterations and the forward or backward address jumps to make on each iteration. The address jump scheme reduces the logic to a set of small accumulators to control the loops and small adders to compute addresses. Innermost loops are usually set to stride over the bit depth of the activations and weights. Outer loops are used to iterate over the bit combinations for the serial dot-product procedure and over the dimensions of the tensors. 
For GEMV, two nested loops are required for both activations and weights. Conv2D operations are programmed to compute one row of the output activation map per job, requiring four nested loops.


\subsubsection{Pipeline Modules}
\label{sec:Pipeline Modules}
Each \MVU{} has modules downstream from the MVP to implement other DNN operations including a multiplier/adder unit, a pooling/ReLU unit, and a quantizer/serializer unit. These modules operate at high-precision. Fixed-point multiplier/adder units (\emph{Scaler} in Figure \ref{fig:Barvinn_top}), compute DNN operations such as batch normalization and quantization scaling as in LSQ \cite{Esser2020LEARNED}. Scalers multiply the MVP output by a 16-bit operand sourced 
from the scaler RAM. In an FPGA, the multiplier is $27\times16$, which aligns with the port widths of on-chip fixed DSP units. An adder that follows adds 32-bit fixed-point bias terms from bias RAM. Scaler and bias RAMs have independent AGUs. The module that follows combines max pooling and ReLU (\emph{Pool/ReLU} in Figure \ref{fig:Barvinn_top}), implemented as a comparator with an internal register. For ReLU, the incoming value is checked against the register initially set to 0. The combined MaxPool/ReLU is implemented by programming  \MVU s to produce data in the sequence needed for a MaxPool window.

The pipeline ends at the quantization/serialization unit (\emph{QuantSer} in Figure \ref{fig:Barvinn_top}). It takes 32-bit fixed-point data from each of the 64 data paths and serializes them into 64 1-bit outputs. It is programmed to set the output bit-depth and the MSB position from the input word. Combined with scaler units, this is used to implement quantization schemes such as LSQ \cite{Esser2020LEARNED}. Serialized outputs of each datapath are grouped into a single 64-bit word that is sent either to the activation RAM of the same \MVU{}, or to a different \MVU{} via an interconnect.

\subsubsection{Interconnect}

\MVU{}s can send data to each other via an interconnect implemented as an 8-way crossbar switch with broadcast capability. A source \MVU{} is programmed to send its output results in a serialized fashion to a given address in the activation memory of a destination \MVU{}(s). At a destination \MVU{}, a fixed-priority arbitration scheme to the write port of the target MVU activation RAM is used. The interconnect is given highest priority, followed by the controller, then lastly the \MVU{} itself. When multiple \MVU{}s attempt to write to the same destination \MVU{}, a fixed priority scheme determines which \MVU{} can write to its memory.

\subsubsection{DNN Mapping}

Each \MVU{} can be assigned to different layers of a DNN, such as convolutions and fully-connected layers. 
Alternatively, a single layer can be split between multiple \MVU{}s with each \MVU{}s processing a subset of the input activations and/or weights. Partial results are forwarded from one \MVU{} to another via the interconnect to process subsequent layers of the network, thus creating an overall processing pipeline through the array. By sending partial results from one \MVU{} to another, subsequent \MVU{}s can begin processing as soon as sufficient data has been received from previous layers. For instance, a \MVU{} processing a $3\times3$ convolution requires only 3 rows of  activations from the previous layer to produce one output row of the layer it is processing. This avoids the need to wait until all outputs from a layer are generated, which reduces latency and idle time. Furthermore, the ability to immediately process partial layer outputs by subsequent \MVU{}s keeps on-chip storage requirements low, since only the partial set of activations required to produce the next layer partial output needs to be stored.

\begin{figure}[!h]
  \centering
  \begin{tikzpicture}

    \draw (0, 0) node[inner sep=0] {\includegraphics[width=8.5cm]{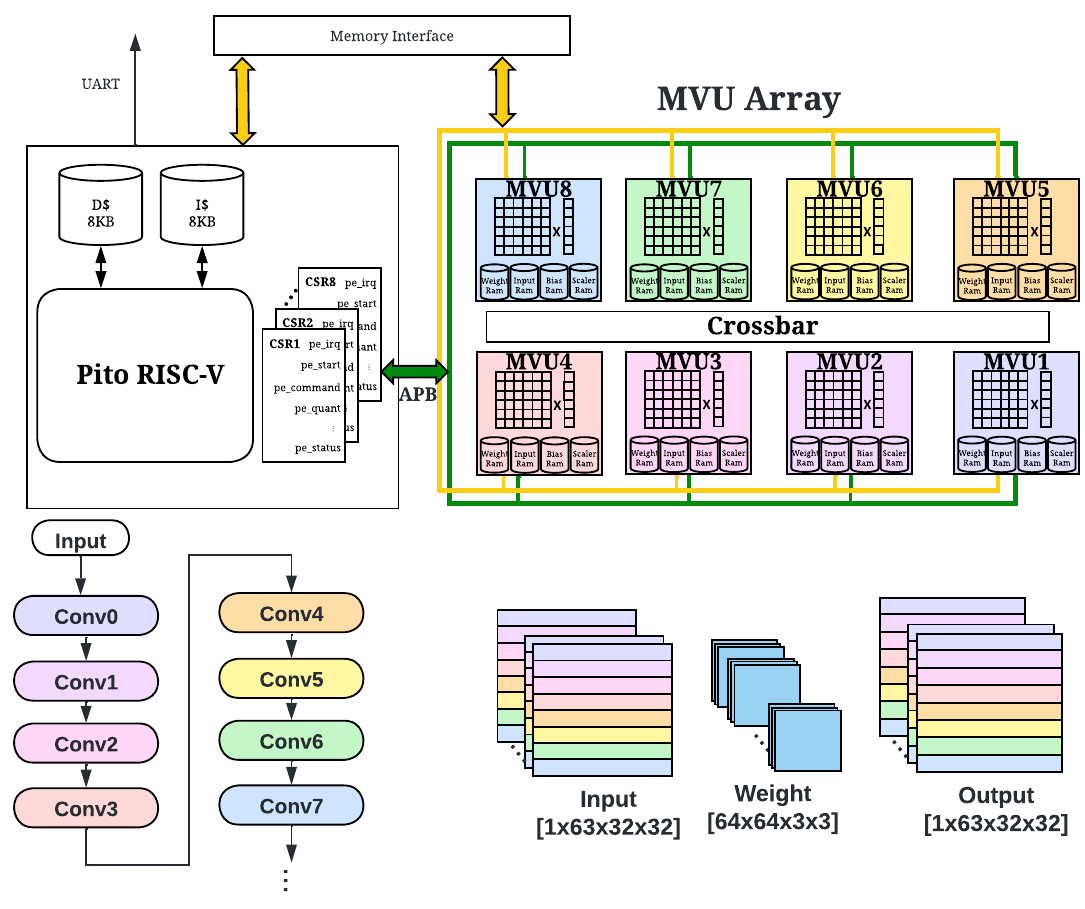}};
    \draw (-2, -3.6) node {a. Pipelined mode};
    \draw (1.5, -3.6) node {b. Distributed mode};

\end{tikzpicture}
    \small
\caption{Execution flow of a DNN on the \MVU{} array in Pipelined (a) and Distributed (b) modes. In Pipelined mode each \MVU{} processes one layer at a time. In distributed mode, the computation of a single layer is distributed among multiple \MVU{}s.}
\label{fig:mvu_laps}
\vspace{-4mm}
\end{figure}

Depending on the performance goal, \BARVINN{} can execute a DNN in either Pipelined mode or Distributed mode.
In Pipelined mode (Figure \ref{fig:mvu_laps}.a), the \MVU{} array can process up to 8 convolutions and fully-connected layers all at once. Each \MVU{} can be configured to use different precisions. In cases where a DNN model contains more than 8 layers, the \MVU{} array can be programmed to process the entire model by dividing it into subsets of up to 8 layers each. Each \MVU{} can be loaded with weights from layers in each subset, either all from the start of processing if there is sufficient weight memory available in each \MVU{} or on-the-fly from external memory if not. Output activations from the last \MVU{} in the chain can also be stored temporarily in off-chip memory and fetched later in the case where the first \MVU{} is still processing data from the current lap. 
In the Distributed mode, to minimize latency, the objective is to process single batch inputs as fast as possible. As can be seen in Figure \ref{fig:mvu_laps}.b, in this mode, the computation of a single layer is broken into 8 independent computation regions. All \MVU{}s will be programmed to share the same set of weights. To make sure an \MVU{} computation is independent of those performed on other  \MVU{}s, the user might need to copy the input regions that are shared between computation units. The programmability of \BARVINN{} allows the user to mix and match these execution modes for different layers and models to achieve highest performance.

\subsection{\Pito{}}

To make use of \MVU{}s for neural networks, a control unit is required. The controller is a \barrel{} RISC-V processor designed to control the 8 \MVU{}s using separate but communicating hardware threads (harts) that each manage their respective \MVU{}s. DNN layers are executed either in distributed or in a pipelined fashion, depending on whether the DNN is compiled to maximize throughput or minimize latency. This design allows \MVU{}s to complete tensor operations independently of each other. The drawback is that it requires 8 microprocessors to execute the 8 programs. We instead amortized the fixed costs of the processor by adopting \barrel{} processing. With a 8-way threaded processor, we may assign one thread to control each of the \MVU{}s. Because every thread comes up for execution only every 8 clock cycles, the five pipeline stages (fetch, decode, execute, data read \& writes and commit) can be completely hidden. Branch prediction units are unnecessary. Since tensor operations can require hundreds of cycles to execute on a \MVU{}, the \barrel{} processor can fully turn over dozens of times in the interim, allowing each thread to issue the next command to its \MVU{} in a few instructions.

%
We adopted a Harvard architecture and divided the instruction and data RAM, 8KB each, and shared between all harts. This gives a 1K word space to store data and instructions to control each \MVU{}. The processor executes instructions following compilation order and without any further scheduling. A hart scheduler provides access to the required resources for the hart at each stage. In the fetch stage, each hart loads instructions from the instruction RAM. The program counter (PC) and register file for each hart is different and the hart scheduler indicates which register should be accessed at a given time. The Decode stage decodes instruction and loads source registers or an immediate operand. Our RISC-V controller is compatible with RV32I RISC-V ISA with minimal support for privilege specification to make Control and Status Registers (CSRs) and Interrupts available to interface with the \MVU{} array. In addition to the base CSRs, we have added 74 \MVU{}-specific CSRs to allow software to control the processing element array. These CSRs control different settings within an MVU such as weight and activation precision, AGU's jump settings, input, weight and output memory address and pipeline module selection as described in \ref{sec:Pipeline Modules}.



\subsection{Code Generator}

\BARVINN{} performs GEMM/GEMV, Convolutions, Maxpooling and activation (ReLU). However, it is up to the user to sequence the operations within a DNN with software. To facilitate this, we developed a code generator that takes a DNN described in ONNX \cite{onnx} and configuration settings (weight/input/output precision), and  generates RISC-V code for each operation. The code generator exports weights to the bit-transposed format described in section \ref{subsec:mem_layout}. Since each \MVU{} works on 64-bit words, the code generator tiles each weight tensor in blocks of 64$\times$64. When this cannot be done (either tensor input channel or output channel is not a multiple of 64), we pad the corresponding tile. Currently, our code generator does not apply graph optimization techniques. Also, for now, our code generator supports Pipelined mode execution. In the following section, we used our code generator to map PyTorch models to micro kernel codes which can then be directly used by \barvinn{}.

\section{Performance Analysis and Results}
\label{sec:perf_analysis}

\subsection{Experimental Setup}
\label{sec:pef_neural_barvinn}
To illustrate the performance of \barvinn{}, we chose the ResNet9 image classifier model for the CIFAR10 dataset. We trained and quantized a ResNet9 model on CIFAR10 using LSQ \cite{Esser2020LEARNED} and used the residual distillation \cite{NEURIPS2020_657b96f0} technique to remove shortcut connections (Plain CNN models). In many image classification DNN models such as ResNet \cite{He_2016_CVPR}, the input to the first layer typically consists of less than 64 channels. Furthermore, due to sensitivity of the first and last layer to information loss, most state-of-the-art compression and quantization methods do not apply optimization on input and output layers \cite{Esser2020LEARNED}, hence keeping these layers untouched and in full precision. We have adopted the same technique to compute first and last layers on the host or on the RISC-V controller.

Table \ref{tab:resnet9_quant} shows the performance of ResNet9 on CIFAR10 in the PyTorch framework. Once we were satisfied with the performance of our quantized model, we exported the trained model to ONNX and then used our code generator.
Table \ref{tab:resnet9_perf} illustrates the per layer computation cost of running ResNet9 on \barvinn{} with 2-bit activations and weights. All convolutions use a padding of 1. As discussed before, we skipped running the first and last layer on \barvinn{} and we kept them in their original format. The overall computation takes 194,688 cycles to complete. 

Our design was written in Verilog and synthesized using Xilinx Vivado 2021.1 
for the Xilinx Alveo U250 accelerator card. 
Synthesis results for the  RISC-V controller, the processing array, and the accelerator are presented in Table~\ref{tab:fpga_impl}. 
Power consumption was estimated using the software tools in Vivado.


\begin{table}[]
\centering

\caption{ResNet9 with different bit precision on CIFAR10}
\begin{tabular}{|l|l|l|l|}
\hline
ResNet9 Model                  & Precision & Accuracy & Size (Bytes)\\ \hline\hline
Original       & Fp32      &     90.8\%     & 19605141\\ \hline
Plain-CNN & Fp32      &       91.1\%   & 18912487\\ \hline
Quantized Plain-CNN  & Int2      & 89.2\% & 1181360\\ \hline
\end{tabular}
\label{tab:resnet9_quant}
\end{table}

\begin{table}[]
\centering

    \caption{ResNet9 layers for CIFAR10 dataset and computation cost.  All layers are quantized to 2-bit for activation and weights, except for the first and last layers.}
    \begin{tabular}{lll|l|l|}
    \hline
    \multicolumn{1}{|l|}{Layer} & \multicolumn{1}{l|}{Input}                & Kernel               & Output               & Cycles \\ \hline\hline
    \multicolumn{1}{|l|}{conv0} & \multicolumn{1}{l|}{{[}3, 32, 32{]}}   & {[}64, 3, 3, 3{]}    & {[}64, 32, 32{]}  & N/A    \\ \hline
    \multicolumn{1}{|l|}{conv1} & \multicolumn{1}{l|}{{[}64, 32, 32{]}}  & {[}64, 64, 3, 3{]}   & {[}64, 32, 32{]}  & 34560  \\ \hline
    \multicolumn{1}{|l|}{conv2} & \multicolumn{1}{l|}{{[}64, 32, 32{]}}  & {[}64, 64, 3, 3{]}   & {[}64, 32, 32{]}  & 34560  \\ \hline
    \multicolumn{1}{|l|}{conv3} & \multicolumn{1}{l|}{{[}64, 32, 32{]}}  & {[}128, 64, 3, 3{]}  & {[}128, 16, 16{]} & 17280  \\ \hline
    \multicolumn{1}{|l|}{conv4} & \multicolumn{1}{l|}{{[}128, 16, 16{]}} & {[}128, 128, 3, 3{]} & {[}128, 8, 8{]}   & 32256  \\ \hline
    \multicolumn{1}{|l|}{conv5} & \multicolumn{1}{l|}{{[}128, 8, 8{]}}   & {[}256, 128, 3, 3{]} & {[}128, 8, 8{]}   & 16128  \\ \hline
    \multicolumn{1}{|l|}{conv6} & \multicolumn{1}{l|}{{[}128, 8, 8{]}}   & {[}256, 256, 3, 3{]} & {[}256, 4, 4{]}   & 27648  \\ \hline
    \multicolumn{1}{|l|}{conv7} & \multicolumn{1}{l|}{{[}256, 4, 4{]}}   & {[}512, 256, 3, 3{]} & {[}256, 4, 4{]}   & 13824  \\ \hline
    \multicolumn{1}{|l|}{conv8} & \multicolumn{1}{l|}{{[}256, 4, 4{]}}   & {[}512, 512, 3, 3{]} & {[}512, 4, 4{]}   & 18432  \\ \hline
    \multicolumn{1}{|l|}{fc}    & \multicolumn{1}{l|}{{[}512, 4, 4{]}}   & {[}10, 512{]}        & {[}10{]}           & N/A    \\ \hline
                                &                                           &                      & Total:               & 194688 \\ \cline{4-5} 
    \end{tabular}
    \label{tab:resnet9_perf}
\vspace{-0.5cm}
\end{table}

\subsection{ Discussion  }

We compared \BARVINN{} with FINN \cite{umuroglu2017finn}, which is a templated Vivado HLS C++ library of common DNN layers. Like \BARVINN{}, FINN can generate hardware for arbitrary precision, but is not software programmable. Hence, once the FINN hardware is generated, the user cannot change the computation data stream. We attempted to compare the performance of \BARVINN{} with FINN using the ResNet9 model we used earlier. However, at the time of writing, FINN supports simple linear topologies and we were not able to get performance metrics for our model. Instead, we used the available CIFAR10-CNV model from the FINN repository that was tuned for the FINN dataflow for our comparison. Table \ref{tab:finn_vs_barvinn} shows the performance of \BARVINN{} and FINN. For this experiment, we used different precisions for weights and activation. For both tools, we used the performance estimation numbers for frames per second (FPS). For FINN, we used the default folding configurations publicly available in FINN-example repository \cite{finn-example}. As illustrated in Table \ref{tab:finn_vs_barvinn}, we provide 7-15 times better throughput
albeit with higher LUT usage. On the other hand, for higher bit precisions, FINN provides a better FPS/LUT, suggesting a scalable solution for bigger models. 

We also compared the performance on a ResNet-50 model. Table \ref{tab:resnet_perf} shows our estimated FPS for \BARVINN{} executing in Pipelined mode along with reported performance for FINN \cite{finn-example} synthesized for the Xilinx U250 and for FILM-QNN \cite{sun2022film} synthesized for the Xilinx ZCU102 FPGA. While FINN has the highest FPS, \BARVINN{} shows the best performance per Watt.
According to the FINN-example repository \cite{finn-example}, a fine-tuned ResNet50 model, requires more than 87\% of Alveo U250 accelerator's resources. This shows the limits of FINN dealing with bigger models. \BARVINN{} requires the same LUT usage regardless of the model size and bit-width.

\begin{table}[]
\caption{Post-synthesis resource utilization of \barvinn{}.}
\begin{tabular}{|l|c|c|c|}
\hline
Resource    & \pito{} RISC-V     & \MVU{} Array       & Overall  \\ \hline\hline
LUT         & 10454    &  190625   &       201079    \\ \hline
BRAM      & 15       &    1312     &      1327      \\ \hline
DSP         &   0       & 512       &      512      \\ \hline
Dynamic Power          &  0.410 W    &   21.066 W  &      21.504 W      \\ \hline
Frequency          &  250 MHz    &   250 MHz  & 250 MHz    \\ \hline
\end{tabular}
\label{tab:fpga_impl}
\end{table}

\begin{table}[]
\centering
\caption{Estimated performance of running CNV model on CIFAR10 on Alveo U250 when different bit precision is used.}
\begin{tabular}{l|c|r|r|r|r|r|}
\cline{2-7}
 & \begin{tabular}[c]{@{}l@{}}Bits\\ (W/A)\end{tabular} & kLUT & \begin{tabular}[c]{@{}l@{}}BRAM\end{tabular} & DSP & FPS & \begin{tabular}[c]{@{}l@{}}FPS/\\ kLUT\end{tabular} \\ \hline 
\multicolumn{1}{|l|}{\multirow{3}{*}{Ours}} & 1/1 & 201.1 (15.0\%) & 1327 & 512 & 61035 & 303.5\\ \cline{2-7} 
\multicolumn{1}{|l|}{} & 1/2 & 201.1 (15.0\%)  & 1327 & 512 & 30517 & 151.7 \\ \cline{2-7} 
\multicolumn{1}{|l|}{} & 2/2 & 201.1 (15.0\%)  & 1327 & 512 & 15258 & 75.8\\ \hline
\multicolumn{1}{|l|}{\multirow{3}{*}{FINN}} & 1/1 & 28.2 (2.1\%) & 150 & 0 & 7716 &  273.6\\ \cline{2-7} 
\multicolumn{1}{|l|}{} & 1/2 & 19.8(1.47\%) & 103 & 0 & 2170 & 109.6\\ \cline{2-7} 
\multicolumn{1}{|l|}{} & 2/2 & 24.3(1.81\%) & 202 & 0 & 2170 & 89.3\\ \hline
\end{tabular}
\label{tab:finn_vs_barvinn}
\end{table}

\begin{table}[]
\centering
\caption{Performance for ResNet-50 model on ImageNet.}
\begin{tabular}{l|c|r|r|r|}
\cline{2-5}
 & \begin{tabular}[c]{@{}l@{}}Bits (W/A)\end{tabular} &  Clock Freq. & FPS & \begin{tabular}[c]{@{}l@{}}FPS/Watt\end{tabular} \\ \hline 
\multicolumn{1}{|l|}{\multirow{1}{*}{Ours}} & 1/2 & 250 MHz & 2296 & 106.8 \\ \cline{1-5} 
\multicolumn{1}{|l|}{\multirow{1}{*}{FINN-R \cite{finn-example}\cite{blott2018finnr}}} & 1/2 & 178 MHz & 2873 & 41.0 \\ \cline{1-5}
\multicolumn{1}{|l|}{\multirow{1}{*}{FILM-QNN \cite{sun2022film}}} & 4(8)/5 & 150 MHz & 109 & 8.4 \\ \cline{1-5} 
\end{tabular}
\label{tab:resnet_perf}
\vspace{-0.2cm}
\end{table}

    
\section{Conclusion}
\label{sec:conclusion}
In this paper, we presented an FPGA-based DNN accelerator that supports arbitrary bit precision computations. We tested the performance of \BARVINN{} over different DNN kernels and models with different bit precision. For model deployment, we developed a code generator tool that takes in a model in ONNX format and generates RISC-V assembly code for the controller. Compared to other low precision accelerators, we provide a programmable solution which makes \BARVINN{} more flexible.  With the programmable \MVU s, the user can run different models regardless of their size. \BARVINN{} allows trading off throughput and latency by running DNN layers either in distributed or in pipeline modes. Unlike other low precision accelerators, our proposed solution offers implementing various trade-offs through software and the end user can control them for each individual layer without FPGA reconfiguration at run time. Compared to programmable accelerators, \BARVINN{} was shown to provide a better throughput per Watt performance. 




\begin{acks}
The authors acknowledge support for this project from the IBM AI Horizons Network, CMC Microsystems, Fonds de Recherche du Quebec–Nature et Technologies (FRQNT), MITACS and from the NSERC COHESA Strategic Research Network.

\end{acks}

\bibliographystyle{ACM-Reference-Format}
\bibliography{refs}




    

\end{document}